# High sensitive quasi freestanding epitaxial graphene gassensor on 6H-SiC

I. Iezhokin[1], P. Offermans[2], S. H. Brongersma[2], A. J. M. Giesbers[1, a)], C. F. J. Flipse[1, b)]

[1] Molecular Materials and Nanosystems, Eindhoven University of Technology, 5600 MB Eindhoven, The Netherlands

[2] Holst Centre/imec, 5656 AE Eindhoven, The Netherlands

Contact information: [a)] A.J.M.Giesbers@tue.nl, [b)] C.F.J.Flipse@tue.nl

We have measured the electrical response to $NO_2$, $N_2$, $NH_3$ and CO for epitaxial graphene and quasi freestanding epitaxial graphene on 6H-SiC substrates. Quasi freestanding epitaxial graphene shows a 6 fold increase in $NO_2$ sensitivity compared to epitaxial graphene. Both samples show a sensitivity better than the experimentally limited 1 ppb. The strong increase in sensitivity of quasi freestanding epitaxial graphene can be explained by a Fermi-energy close to the Dirac Point leading to a strongly surface doping dependent sample resistance. Both sensors show a negligible sensitivity to $N_2$, $NH_3$ and CO.

Graphene is an ideal candidate for gas sensing due to its two-dimensional nature, consisting mainly of a surface with a high mobility and a doping dependent resistance. Therefore, recently several graphene allotropes such as exfoliated graphene flakes,[1-3] CVD graphene,[4-5] chemically reduced graphene oxide,[6-8] epitaxial graphene,[9, 10] nanostructured graphene[11] and graphene foam[12] were successfully used as proof-of-principle gas sensor for the most common environmental pollutants such as nitrogen oxides, carbon oxides and sulfur oxides. Detection down to a single molecule was claimed[1] with charge transfer being the main mechanism behind the sensing, i.e. absorbed molecules on top of the graphene either donate or accept an electron and thereby change the graphene resistance. Graphene seems to be most promising for sensing nitrogen dioxide ($NO_2$) where ppt-level sensitivity was claimed for CVD graphene under constant UV illumination.[4] Without UV illumination, UHV-grown epitaxial graphene was shown to give the highest sensitivity (500 ppb) and selectivity to $NO_2$.[9] This is comparable to current state-of-the-art proof-of-principle solid state sensors such as metal oxides[13] and carbon nanotubes.[14] A major advantage of epitaxially grown graphene over the other allotropes is its clean fabrication which avoids processing pollutants that can influence the sensing, such as polymethylmethacrylaat (PMMA),[15] and allows easy up-scaling. The quality of epitaxial graphene was shown to increase considerably by growing at atmospheric pressures[16] and by hydrogen intercalation[17] of graphene on SiC, compared to UHV-grown graphene. Here we use both of these growth methods to fabricate high sensitive graphene gas sensors and characterize them for $NO_2$, $NH_3$, $N_2$, and CO. We show an increased sensitivity of more than two orders of magnitude compared to vacuum-grown epitaxial graphene.[9] The relative resistance changes over 1%, with a signal-to-noise ratio of $10^{-4}$, for hydrogen-intercalated epitaxial graphene within 2 minutes of exposure to 1 ppb $NO_2$. The increased quality for quasi-freestanding graphene is qualitatively explained by a shift of the Fermi energy close to the Dirac point.

We use epitaxial graphene (eG) and quasi-freestanding epitaxial graphene (QFeG) samples for gas sensing. Figure 1(a) shows a typical sensor device layout. The graphene on top of the SiC substrate is electrically connected by silver epoxy and changes in the graphene resistance are used to detect gas molecules on the sample surface. Directly below the sample we mounted a resistive Pt heater to anneal the sample up to 150 $^o$C to its pristine state prior to each measurement. Both eG and QFeG samples are grown by thermal decomposition on the silicon (0001) side of a 4×4 mm$^2$ insulating 6H-SiC wafer piece from II-VI Inc. For the growth of eG we followed the procedure described by Emtsev et al.[16] The result is a decoupled layer of graphene supported by a buffer layer, which consists of a carbon layer strongly bonded to the substrate (see top right inset Fig. 1(b)).[18] The main panel in Fig. 1(b) shows an atomic force micrograph of the topology after growth. Clearly visible is the terrace structure originating from a slight miss cut of the SiC substrate. The terraces are typically 1-2 μm in width and have a step-height of about 10 nm (see inset Fig. 1(b)). In the phase image (Fig. 1(c)), recorded simultaneously with the topology, we can see that the terraces consist mainly of single layer (1L) graphene with a narrow region of bilayer (2L) graphene on the terrace edges. This layer sequence was confirmed by Raman microscopy (see supplementary information Fig. S1).

For the QFeG samples we used a similar procedure as developed by Riedl et al.[17] First we grow a buffer layer by thermal decomposition of carbon at 1450 °C under an atmospheric flow (0.1 slm) of argon (grade 5.0) for 15 minutes. In a second step the buffer layer is intercalated by hydrogen by annealing for 75 minutes at 550 °C in a 930 mbar hydrogen flow (0.9 slm, grade 5.0). The result is a graphene layer which is decoupled from the hydrogen passivated SiC substrate (see top right inset Fig. 1(d)). The topology of the QFeG is shown in the main panel of Fig. 1(d) together with the phase image in Fig. 1(e). The terraces are narrower (0.5-1 μm) than the terraces in eG, however, their single layer coverage is much larger. Only a few narrow uncovered patches (dark areas in phase image) are visible on the terrace edges. Bilayer regions are almost completely absent in our QFeG samples. This is confirmed by the unchanged Raman spectra over the entire surface (see supplementary information Fig. S1). Figure 1(f) shows the single layer Raman spectra for both the eG and the QFeG. Clearly visible is the increased quality of the QFeG by the narrow G and 2D peak.[19] The QFeG, however, shows a clear D-peak related to defects indicating that the buffer layer grows with more defects than the first graphene layer. Room temperature Hall resistance measurements are used to determine the carrier concentration, $n$, and mobility, $\mu$, of the samples leading to $n_{eG} = 2 \cdot 10^{13}$ cm$^{-2}$ and $n_{QFeG} = 1.5 \cdot 10^{12}$ cm$^{-2}$ with $\mu_{eG}$ = 200 cm$^2$/Vs and $\mu_{QFeG}$ = 800 cm$^2$/Vs at room temperature, respectively. This corroborates the improved quality of QFeG over that of eG.[20]

To characterize and compare eG and QFeG as gas sensors we measured both samples simultaneously in a flow chamber supplied with a constant 250 sccm gas flow at atmospheric pressure and room temperature (25 °C). Nitrogen (99.9995%, 5.5 N purity) is used as a carrier gas. $NO_2$ and $NH_3$ were supplied from a certified permeation tube (KIN-TEK) using a permeation oven (MCZ) and CO was supplied from a gas cylinder (Praixair 99.9995%, 5.5 N purity). The test gases were diluted by the carrier gas using a gas calibration system (MCZ). The resistance of the device was measured by a four probe compensation method using a Keitley SourceMeter (K2612a). A constant current of 1 mA was applied to the sensor and the relative resistance change was recorded during gas exposure. Prior to the measurements, the sample was annealed in vacuum at a temperature of 150 °C using the resistive heater for 30 minutes to get the sample in its pristine condition. Subsequently, a pure $N_2$ gas flow was added to verify the inertness of graphene to nitrogen. No electrical response was detected within the measurement noise (signal to noise ratio at base line is $\Delta R/R=10^{-4}$) and the resistance was stable over several hours (see first 30 min. $N_2$ lines in Fig. 2(a) and (b)).

After initialization the sensor gas is added to the flow and the relative resistance change of the two samples is monitored. Figure 2 (a) and (b) shows the respective results for the eG and the QFeG sample for the different sensor gases. The response or relative change in resistance ($R-R_0$)/$R_0$, where $R_0$ is the resistance prior to exposure, of the eG sample shows a strong initial increase of almost 30% in 1 minute for 100 ppb $NO_2$. Within 30 minutes the response approaches a steady-state with a change over 50%, suggesting equilibrium between gas absorption and desorption. After 30 minutes of exposure the sensor gas supply is turned off and the sample is left in a $N_2$ flow for 30 minutes. As the adsorption rate is set to zero, by stopping the sensor gas supply, desorption of the sensor gas is clearly visible in the decreasing responce. However even after several hours of waiting the initial resistance is not reached (not shown). To regain the initial resistance we heat our sample up to 150 °C for 30 minutes. The strong responce directly after switching on the heater is caused by the strongly temperature dependent phonon scattering in epitaxial graphene,[21, 22] and is immediately followed by a decrease due to gas desorption. Upon cooling the phonon scattering contribution disappears, visible by the strong response drop after the heater is switched off. The same temperature effect can be seen clearly in the response in inert $N_2$ (black line Fig. 1(a)). In addition to $NO_2$ we measured the relative resistance change to 300 ppm $NH_3$ and 3000 ppm CO (both supply-limited concentrations) and observed no response. We measured the response of all gasses both parallel and perpendicular to the substrate terraces (not shown). The resistance perpendicular to the terraces is always higher due to the additional resistance created by the terrace steps.[23] A similar, less pronounced, resistance asymmetry is also observed for the QFeG sample. Figure 2(b) shows high resistance channel for the QFeG sample with the same exposure sequence as described above for the eG sample. Directly after switching on the $NO_2$ sensor gas flow (40 ppb) we observe a dramatic response of over 300% for $NO_2$ in the first 3 minutes. After this strong initial increase the response reaches a maximum and goes down for longer exposure times. After the sensor gas is switched off the response increases again until it reaches a maximum and subsequently goes down (not shown here). This process is accelerated by thermally heating the sample. In the sequence shown in Fig. 2(b) the heater is switched on at 60 minutes for quick recovery, similar to the eG sample. The response nevertheless shows the same behavior as described for the non-heated case with a peak a short time after the start of the heating procedure. The resistance in QFeG is much less affected by phonon scattering and therefore a strong initial increase of the resistance due to the temperature change is absent. There is no response for $NH_3$ on QFeG and only a 0.5% change for 3000 ppm CO in 30 min. on QFeG. This is consistent with first-principles calculations on charge transfer of molecules on a graphene surface.[24, 25] For $NO_2$ the LUMO level is only 0.3 eV below the Dirac point and the HOMO level ~ 1 eV, leading to a large charge transfer to the molecule (0.1$e$), making $NO_2$ a strong acceptor. In the case of CO, the calculated charge transfer is towards the graphene (donor) and relatively small (0.01$e$) as the HOMO level is far away (5 eV) from the Dirac point and the LUMO level does not participate in the charge transfer due to symmetry considerations. The small charge transfer is in agreement with a very weak response in the experiment. However, in the experiment CO acts as a weak acceptor. $NH_3$ is also calculated to be a weak donor (0.03$e$). Due to the 10 fold lower maximum concentration compared to CO in the experiment no response to $NH_3$ is expected, consistent with the observations. The charge transfer for $N_2$ is probably even less compared to CO and $NH_3$ as it is a closed shell molecule and consequently the HOMO and LUMO levels are even further away from the Dirac point.

Figure 2 (c) and (d) show the response of both samples for different $NO_2$ concentrations between 1 and 40 ppb. The same sequence of exposure, idle, heating and idle as described above is used here. In general, the lower the concentration the longer it takes to achieve the same response. For the QFeG sample (Fig. 2(d)) we observe that if the response did not reach a maximum during exposure it goes down in the 'gas off' phase; in contrast, if it did reach a maximum it goes up as described for Fig. 2(b) (40 ppb case). Similarly, for the subsequent heating step the response initially goes up if a peak was reached in the 'gas on' phase and directly goes down if no maximum was reached. The QFeG showed a response of almost 40% in 30 min exposure at a $NO_2$ concentration of only 1 ppb. This response is a factor of 6 higher than our eG sample (same holds for the higher concentrations), showing the strongly improved performance of QFeG as a gas sensor. The sensitivity of both our devices outperforms previous epitaxial graphene based gas sensors[9, 10] and is comparable to the sensitivity of current state of the art solid state gas sensors.[13, 14] In addition to the high sensitivity, our QFeG samples show response times for a 1% change of 2 minutes at 1 ppb and 18 seconds at 10 ppb, comparable to carbon nanotube based sensors.[14]

To explain the resistance behavior of our two graphene samples we use a simple model as illustrated in Fig. 3. In their pristine condition the main important difference between eG and QFeG is the position of the Fermi-energy (Fig. 3(a)). The eG samples are highly electron doped ($n_{eG} = 2 \cdot 10^{13}$ cm$^{-2}$) due to the presence of the buffer layer.[16] After hydrogen intercalation in QFeG the doping is mostly removed ($n_{QFeG} = 1.5 \cdot 10^{12}$ cm$^{-2}$) thereby positioning the Fermi-level close to the Dirac point or charge neutral point (CNP).[17] Based on the data by Waldmann *et al.*[26] we can translate this initial condition to a schematic illustration of the resistance versus energy (Fig. 3(b) solid dark blue for eG and solid red for QFeG). As the density of states in graphene is linearly proportional to the energy, $E \propto n$, the x-axis can also be viewed as an induced carrier concentration. The resistance peak of the QFeG is drawn slightly narrower representing its improved quality over eG in analogy to annealed graphene flakes.[27] For the eG sample, the Fermi-energy is in the tail of the resistance peak which has is maximum at the CNP. Adding the sensor gas molecules to the surface of eG causes hole doping of the system leading to a shift of the resistance peak closer to the Fermi-energy (indicated by the dark blue arrow).[10] At the Fermi-energy this leads to an increase in the resistance. Assuming a constant increase in time of the sensor gas molecules on the surface, the resistance changes as illustrated in the inset of Fig. 3(b) (dark blue dashed line). This is indeed close to what we observe in the experiment. For the QFeG sample, the resistance peak is situated much closer to the Fermi-energy (solid red line Fig. 3(b)). A change in the resistance peak position due to surface doping leads to a much larger resistance change at the Fermi-energy compared to the same shift in eG. For large shifts we can see that the resistance peak passes the Fermi-energy and changes the system from electron doped to hole doped. The resistance *versus* time plot shows a peak in this case (inset Fig. 3(b)), which is indeed what we also observed in the experiment (Fig. 2(b) and (d)). Removing the molecular doping, e.g. by heating, shifts the resistance peak back to its initial position, thereby passing the Fermi-energy, leading to the peak observed during the annealing step.

To conclude, we observed a six-fold increase in sensitivity for quasi-freestanding epitaxial graphene compared to epitaxial graphene. This increase can be understood by a strong reduction of background doping in QFeG, positioning the Fermi-energy close to the CNP where the graphene resistance depends strongly on changes in surface doping. Both samples showed an extremely high sensitivity, < 1 ppb (supply limit), and a fast response time to $NO_2$ gas. No changes were observed for pure $N_2$, $NH_3$ and CO as is expected due to the position of the HOMO and LUMO levels, which are far away from the Dirac point, resulting in a very small or negligible charge transfer between molecule and graphene.

We would like to thank Th. Seyller for support on the growth of epitaxial graphene. A.J.M.G. acknowledges financial support by the Dutch Organization for Scientific Research (NWO) under Project No. 11447.

[1] F. Schedin, A. K. Geim, S. V. Morozov, E. W. Hill, P. Blake, M. I. Katsnelson, K. S. Novoselov, Nature Matt. **6**, 652 (2007).

[2] Y. Dan, Y. Lu, N. J. Kybert, Z. Luo, and A. T. C. Johnson, Nano Lett. **9**, 1472 (2009).

[3] S. Rumyantsev, G. Liu, M. S. Shur, R. A. Potyrailo and A. A. Balandin, Nano Lett. **12**, 2294 (2012).

[4] G. Chen, T. M. Paronyan, and A. R. Harutyunyana, Appl. Phys. Lett. **101**, 053119 (2012).

[5] A. K. Singh, M. A. Uddin, J. T. Tolson, H. Maire-Afeli, N. Sbrockey, G. S. Tompa, M. G. Spencer, T. Vogt, T. S. Sudarshan, and G. Koley, Appl. Phys. Lett. **102**, 043101 (2013).

[6] G. Lu, S. Park, K. Yu, R. S. Ruoff, L. E. Ocola, D. Rosenmann, J. Chen, ACS Nano **5**, 1154 (2011).

[7] G. Lu, L. E. Ocola, J. Chen, Appl. Phys. Lett. **94**, 083111 (2009).

[8] I-S. Kang, H-M. So, G-S. Bang, J-H. Kwak, J-O. Lee, and C. W. Ahn, Appl. Phys. Lett. **101**, 123504 (2012).

[9] M. W. K. Nomani, R. Shishir, M. Qazi, D. Diwan, V. B. Shields, M. G. Spencer, G. S. Tompa, N. M. Sbrockey, G. Koley, Sens. Actuators B: Chem. **150**, 301 (2010).

[10] R. Pearce, T. Iakimov, M. Andersson, L. Hultman, A. L. Spetz, R. Yakimova, Sens. Actuators B: Chem **155**, 451 (2011).

[11] R. K. Paul, S. Badhulika, N. M. Saucedo, and A. Mulchandani, Analytical Chem. **84**, 8171 (2012).

[12] F. Yavari, Z. Chen, A. V. Thomas, W. Ren, H-M. Cheng, and N. Koratkar, Sci. Rep. **1**, 166 (2011).

[13] K. Wetchakun, T. Samerjai, N. Tamaekong, C. Liewhiran, C. Siriwong, V. Kruefu, A. Wisitsoraat, A. Tuantranont, S. Phanichphant, Sens. Actuators B: Chem. **160**, 580 (2011).

[14] T. Zhang, S. Mubeen, N. V. Myung, and M. A. Deshusses, Nanotech. **19**, 332001 (2008).

[15] Y. Dan, Y. Lu, N. J. Kybert, Z. Luo, and A. T. C. Johnson, Nano Lett. **9**, 1472 (2009).

[16] K. V. Emtsev, A. Bostwick, K. Horn, J. Jobst, G. L. Kellogg, L. Ley, J. L. McChesney, T. Ohta, S. A. Reshanov, J. Rohrl, E. Rotenberg, A. K. Schmid, D. Waldmann, H. B. Weber, and Th. Seyller, Nature Mat. **8**, 203 (2009).

[17] C. Riedl, C. Coletti, T. Iwasaki, A. A. Zakharov, and U. Starke, Phys. Rev. Lett. **103**, 246804 (2009).

[18]C. Riedl, C. Coletti, and U. Starke, J. Phys. D: Appl. Phys. **43**, 374009 (2010).

[19]A. C. Ferrari, J. C. Meyer, V. Scardaci, C. Casiraghi, M. Lazzeri, F. Mauri, S. Piscanec, D. Jiang, K. S. Novoselov, S. Roth, and A. K. Geim, Phys. Rev. Lett. **97**, 187401 (2006).

[20]F. Speck, J. Jobst, F. Fromm, M. Ostler, D. Waldmann, M. Mundhausen, H. B. Weber, and Th. Seyller, Appl. Phys. Lett. **99**, 122106 (2011).

[21]S. Tanabe, Y. Sekine, H. Kageshima, M. Nagase, and H. Hibino, Phys. Rev. B **84**, 115458 (2011).

[22]C. Yu, J. Li, Q. B. Liu, S. B. Dun, Z. Z. He, X. W. Zhang, S. J. Cai, and Z. H. Feng, Appl. Phys. Lett. **102**, 013107 (2013).

[23]T. Low, V. Perebeinos, J. Tersoff, and Ph. Avouris, Phys. Rev. Lett. **108**, 096601 (2012).

[24]O. Leenaerts, B. Partoens, and F. M. Peeters, Phys. Rev. B **77**, 125416 (2008).

[25]T.O. Wehling, M.I. Katsnelson, A.I. Lichtenstein, Chem. Phys. Lett. **476**, 125 (2009).

[26]D. Waldmann, J. Jobst, F. Speck, Th. Seyller, M. Krieger, H. B. Weber, Nature Matt. **10**, 357 (2011).

[27]Z. Cheng, Q. Zhou, C. Wang, Q. Li, C. Wang, Y. Fang, Nano Lett. **11**, 767 (2011).

**Figures**

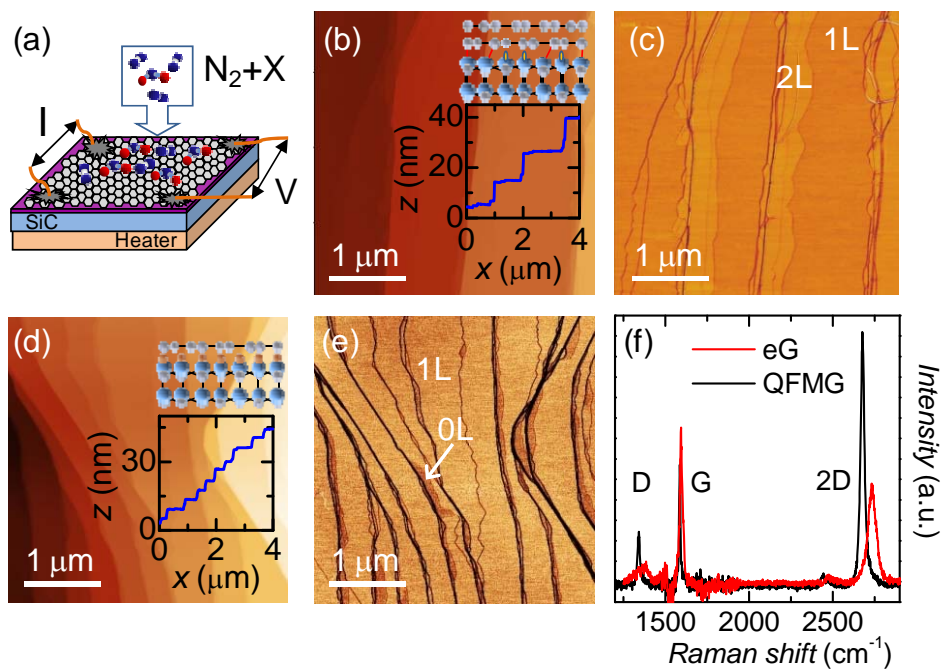

FIG. 1. (a) Schematic illustration of the sensing experiment where the sensing gas X diluted in Nitrogen carrier gas flows over a graphene-on-SiC sample while the resistance change of the graphene is measured. (b) Atomic force micrograph of an epitaxial graphene sample illustrating the terraces (see also inset). The inset illustration schematically shows a cross-section of the sample with an epitaxial graphene layer on top of the buffer layer and the SiC-substrate. (c) AFM Phase image showing the single layer graphene areas (dark) and double layer graphene areas (bright) on top of the terraces. (d) Atomic force micrograph of an H-intercalated graphene sample with its terrace steps shown in the inset. The inset illustration shows a schematic cross-section of the sample with a quasi freestanding graphene layer on top of the hydrogen passivated SiC substrate. (e) AFM Phase image showing the full coverage of freestanding single layer graphene with a few small patches uncovered (dark). (f) Raman spectra of epitaxial and quasi freestanding graphene.

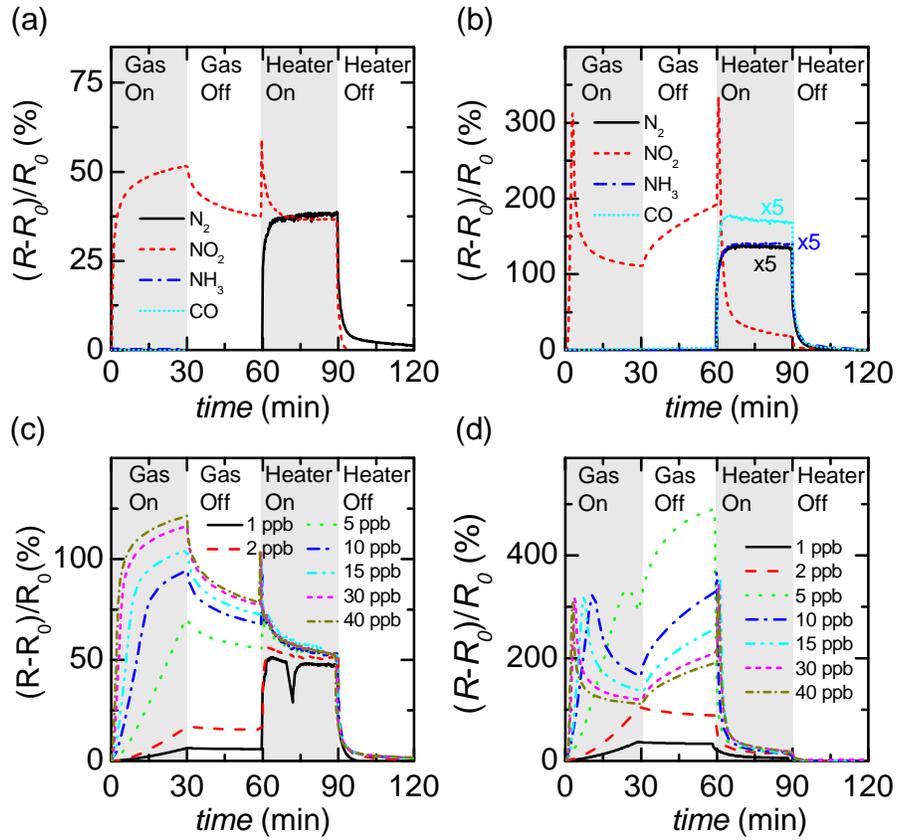

FIG. 2. Resistance change of (a) epitaxial graphene (eG) and (b) intercalated epitaxial graphene (QFeG) exposed to pure $N_2$, 100 ppb $NO_2$ (for eG), 40 ppb $NO_2$ (for QFeG), 300 ppm $NH_3$ and 3000 ppm CO. During anneal the sample was heated to 150 $^{\circ}$C. Sensitivity of (c) eG and (d) QFeG to various concentrations of $NO_2$.

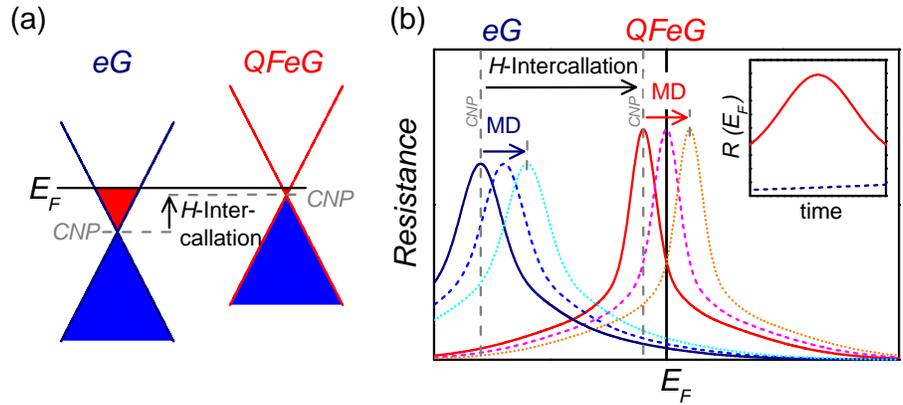

FIG. 3. (a) Schematic illustration of the Fermi-energy shift due to hydrogen intercalation of epitaxial graphene. (b) Schematic illustration of the resistance peak at the charge neutral point (CNP) as a function of energy showing the large CNP shift towards $E_F$ due to hydrogen intercalation and the changes due to molecular doping (MD). Inset: Resistance at the Fermi-energy as a function of exposure time, illustrating the expected resistance change for eG and QFeG due to the molecular doping illustrated in the main panel.

**Supplementary information**

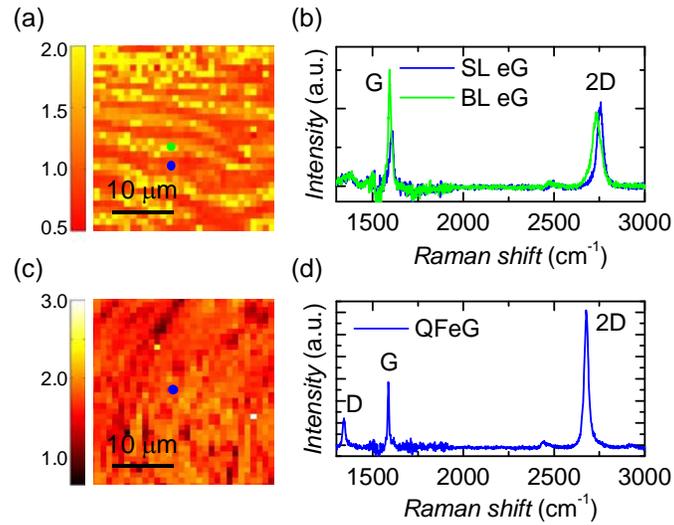

FIG. S1. (a) Raman map of the 2D/G peak intensity ratio showing the distribution of single layer (SL) and Bi-layer (BL) graphene on epitaxial graphene, with SL on the terrace centre and bi-layer on its edges as shown in the AFM images of Fig. 1 in the main text. (b) Raman spectra taken at the dots in (a) with in blue the spectra for SL graphene and in green for BL graphene. (c) Raman 2D/G peak intensity map for quasi freestanding epitaxial graphene displaying full single layer coverage. (d) Raman spectrum of quasi freestanding epitaxial graphene taken at the blue dot in (c).